\documentclass[nanomaterials,article,accept,pdftex,moreauthors]{mdpi} 
\firstpage{1} 
\pubvolume{12}
\issuenum{1}
\articlenumber{3159}
\pubyear{2022}
\copyrightyear{2022}
\externaleditor{Academic Editors:Jean-Marie Nedelec and Jordi Sort}
\datereceived{24 August 2022} 
\dateaccepted{7 September 2022} 
\datepublished{12 September 2022} 
\hreflink{https://doi.org/nano12183159} 
\Title{Dependence of Exchange Bias on Interparticle Interactions in Co/CoO Core/shell Nanostructures}

\TitleCitation{Dependence of Exchange Bias on Interparticle Interactions in Co/CoO Core/shell Nanostructures}

\Author{S. Goswami $^{1}$, P. Gupta $^{2}$\orcidA{}, S. Nayak $^{2,}$, S.Bedanta $^{2,3}$\orcidB{}, {\`{O}}. Iglesias $^{4,}$\orcidC{},  M. Chakraborty $^{1}$ and D. De $^{1,5,*}$\orcidD{}}

\address{
$^{1}$ \quad Material Science Research Lab, The Neotia University, Sarisa, D.H. Road, 24 Pgs (South) West Bengal 743368, India; e-mail@e-mail.com\\
$^{2}$ \quad Laboratory for Nanomagnetism and Magnetic Materials (LNMM), School of Physical Sciences, National Institute of Science Education and Research (NISER), HBNI, Jatni 752050, India; e-mail@e-mail.com\\
$^{3}$ \quad Center for Interdisciplinary Sciences, National Institute of Science Education and Research, HBNI, Bhubaneswar, Odisha 752050, India; e-mail@e-mail.com\\
$^{4}$ \quad Dpt. Física de la Matèria Condensada and IN$^2$UB, Facultat de Física, Universitat de Barcelona, Av. Diagonal 647, 08028 Barcelona, Spain; oscariglesias@ub.edu\\
$^{5}$ \quad Sukumar Sengupta Mahavidyalaya, State Highway 7, Keshpur, Paschim Medinipur 721150, West Bengal, India; mail@mail.com.
}
\corres{Correspondence: sbedanta@niser.ac.in (S.B.); oscariglesias@ub.edu (Ò.I.); manashi.chakraborty@tnu.in (M.C.);
debajyoti.de@tnu.in (D.D.)
}

\abstract{This article reports dependence of exchange bias (EB) effect  on interparticle interactions in nanocrystalline Co/CoO core/shell structures, synthesized using conventional sol-gel technique. Analysis via powder X-Ray diffraction (PXRD) studies and transmission electron microscope (TEM) images confirm absence of crystalline phases other than core-shell Co-CoO with average particle size $\approx$18 nm. 
Volume fraction ($\varphi$) is varied (from 20\% to 1\%) by introduction of stoichiometric amount of non-magnetic amorphous silica matrix (SiO$_2$) which leads to a change in interparticle separation/interaction. 
The influence of exchange and dipolar interactions on the EB effect, caused by the variation in interparticle interaction/separation is studied for a series of Co/CoO core/shell nanoparticle systems. 
Studies of thermal variation of magnetization ($M- T$) and magnetic hysteresis
loops ($M- H$) for the series point towards strong dependence of magnetic properties on dipolar
interaction in concentrated assemblies whereas individual nanoparticle response is dominant in
isolated nanoparticle systems. 
The analysis of the EB effect reveals a monotonic increase of coercivity ($H_C$) and EB field ($H_E$) with increasing volume fraction. When the nanoparticles are close enough and the interparticle interaction is significant, collective behavior leads to an increase in the effective antiferromagnetic (AFM) CoO shell thickness which results in high $H_C$, $H_E$. Moreover, in concentrated
assemblies, the dipolar field superposes to the local exchange field and enhances the EB effect
contributing as an additional source of unidirectional anisotropy.
}


\begin{document}

\section{Introduction}

Nanoscience and nanotechnology fundamentally emerge from manipulation of matter at nanoscale and the curiosity to understand the interactions of matter at the atomic level. In nanoparticle assemblies, parameters such as size \cite{He}, surface structure \cite{Goswami}, shape \cite{Auvinen}, agglomeration \cite{Ashraf} or interparticle interactions \cite{Papaefthymiou} often influence their properties. At the same time, they lead to the emergence of enriched physico-chemical properties, which distinguish  them from their bulk counterparts. Among different classes of nanomaterials, core/shell structures are fundamentally interesting because of carrying two different physico-chemical properties in one single particle at the nanoscale \cite{De2016}. Though primarily, core/shell structures were synthesized to protect and stabilize the metallic core \cite{chaudhuri}, advances in materials fabrication and synthesis have made core/shell structures potential candidates for a myriad of new applications including targeted drug delivery \cite{Schartl}, biomedical sensors \cite{Al-Ogaidi}, enhanced electronic properties \cite{Cha} or EB effect \cite{De2016}. 
If the core and shell are composed of two materials with different magnetic order, the interfacial region will experience a structural modification due to differences in the crystalline structures of both regions as well as 
a competition between the different magnetic orders favored at the core and shell.
This leads to the phenomenon known as exchange bias effect \cite{Feygenson} that was first reported by Meiklejhon and Bean as originating via unidirectional exchange anisotropy in Co-CoO (ferromagnetic (FM) - antiferromagnetic (AFM)) particles \cite{Meiklejhon}. Since then, this has been intensely studied in many magnetically coupled systems such as FM/FiM (ferrimagnetic) \cite{Vasilakaki}, AFM/FiM \cite{Alvarez}, AFM/SG (spin-glass) \cite{Sahoo} or FiM/SG \cite{Wang}.
 
Despite intensive experimental research in the field, there are phenomena like spontaneous exchange bias \cite{SKGiri}, EB in alloys and compounds \cite{Giri2011,Nogues}, EB in single phase magnetically inhomogeneous materials \cite{Goswami-JALCOM} or EB in thin films \cite{Bedanta3,Bedanta4} that are still drawing attention because of the urge to understand new fundamental physics as well as for the wide range of potential applications in recording media to overcome the superparamagnetic limit \cite{Skumryev}, field sensors \cite{Sharma}, read heads \cite{Nogues1999}, giant magnetoresistance (GMR) based devices \cite{Aktas} and many more. 
In this regard, tuning EB related properties by controlling variation of size \cite{Huang} and thickness of core and shell \cite{Feygenson}, interparticle interactions \cite{Sampad1} and microscopic structure of the interface of any magnetically inhomogeneous system \cite{Goswami-JALCOM} might add significant value in several application oriented phenomena. 
Attempts have been made to understand the underlying physics of EB mechanism when the variation of shape \cite{Dimitriadis}, size \cite{Alvarez}, surface composition \cite{Obaidat}, core to shell diameter ratio come into play \cite{De2016}, in a core/shell nanostructure. Recently, we have reported correlating experimental findings and atomistic Monte Carlo (MC) simulations showing that the variation of core and shell thickness of Co-Co$_3$O$_4$ nanostructure leads to systematic changes in the EB effect \cite{De2016}. 

Via controlled oxidation on the surface of transition metal nanoparticles, a shell of metal oxide (generally AFM/FiM in nature) may be formed to prepare a metal/metal oxide core/shell structures \cite{GIRI,Zhang}. Co/CoO is the most studied core-shell nanostructure because of its large interface energy with high EB field ($H_E\approx$1000 Oe) compared to others \cite{Gonzalez,Feygenson}. 
Besides potential technological applications of Co-based nanoparticles in information storage, magnetic fluids, catalysis etc., low crystal anisotropy of Co is favorable for FM/AFM Co/CoO as a model system for EB studies \cite{Simeonidis}. Additionally, because of high AFM N$\acute{e}$el temperature ($T_N$$\approx285$ K) of CoO, followed by wide temperature range of EB effect \cite{Tracy}, different nanostructures of the same has been revisited by researchers to understand different phenomenological models related to EB
In particular, different studies have reported how the shell thickness \cite{Tracy}, the degree of oxidation of the shell \cite{Iglesias2}, the core to shell diameter ratio \cite{Feygenson}, and the degree of dilution within non-magnetic matrix \cite{Nogues2006} affect EB in Co/CoO nanoparticles. 

When particles are in close proximity, the magnetic properties of the nanoparticle assembly are mainly governed by exchange interactions between the surfaces in contact. 
Instead, long range dipolar interactions between the macroscopic magnetic moments of the individual particles can be relevant over a wide range of interparticle separations. As a consequence, it is expected that the thermal and field dependence of the magnetization of the assembly may be very different from that of an individual nanoparticle, giving rise to a variety of behaviors such as superparamagnetism \cite{Bean,Bedanta}, superspin-glass \cite{Bedanta1}, and superferromagnetism \cite{Bedanta,Bedanta2} among others. Some progress has been made in recent studies of frozen ferrofluids \cite{Luo}, granular nanoparticles \cite{Londono} and diluted magnetic systems \cite{Vestal}.

However, to the best of our knowledge, a systematic study to understand the effect of variation of interparticle interactions on EB mechanism in core/shell nanostructure keeping the core and shell diameters fixed, has not been yet reported. In the present article, we aim to study how interparticle interactions among core/shell nanoparticles can influence the phenomenology of EB.
Interparticle interactions may be varied by changing the volume fraction ($\varphi$), but a high level of dilution is required for a comparative study of their effects on the magnetic properties \cite{Papaefthymiou}. Herein, we have performed a systematic and detailed study of the EB effect by changing the interparticle separation of Co/CoO core/shell nanostructures in seven different batches. Experimental findings point that collective magnetization is hindered with the increase in separation of particles via decreasing volume fraction, which leads to a monotonic reduction of coercivity ($H_C$) and EB field ($H_E$).

\section{Experimental}
 
Core/shell nanocrystalline Co/CoO is derived via controlled oxidation-reduction from nanocrystalline Co which is synthesized using conventional sol-gel technique \cite{Goswami}. To begin with, Co metal powder (Aldrich, 99.99\% pure) is dissolved in minimum quantity of 37\% nitric acid and vigorously stirred in a magnetic rotor for $12$ h until the solution become transparent. Stoichiometric amount of citric acid is added in the solution and homogenized for $6$ h to obtain transparent reddish solution. This ensures that all the metal ions are mixed at the atomic scale \cite{De}. The solution is dried very slowly at room temperature for few days. To increase the evaporation rate, solution is kept inside a vacuum oven at 50$^{\circ}$ C for few days. After the solution has transformed into a gel like state, it is heated at 100$^{\circ}$ C to form a cake. This cake is then ground and heated at 600$^{\circ}$ C for 6 h in presence of continuous flow of of Ar-H$_2$ gas (95\% Ar and 5\% H$_2$). Thus Co nanoparticles are obtained. Now, as oxidation will start from the surface, oxide shell is created via controlled oxidation at ambient temperature and the procedure is standardized after repetitive trials. The as-synthesized Co nanoparticle is then heated in open air at 200$^{\circ}$ C for 6 min to form an oxide shell over Co nanoparticle core. The shell consists of both CoO and Co$_3$O$_4$ phases (as evident from XRD pattern describe later). To synthesize desired Co-CoO, this as-synthesized core/shell sample is annealed at 250$^{\circ}$ C in a continuous flow of the same reducing gas (95\% Ar and 5\% H$_2$) for 1 hour \cite{Zhang}. This reduces excess oxygen from Co$_3$O$_4$ and retains only stable CoO shell on the surface of Co core. Thus finally Co/CoO core/shell nanostructure is obtained. 

Variation of interparticle interactions is introduced via incorporation of non-magnetic amorphous silica matrix (SiO$_2$) into Co/CoO core/shell nanoparticles. Stoichiometric amount of SiO$_2$ is added to the as-synthesized sample and homogenized. This reduces the volume fraction and thus increase the interparticle separation leading to the simultaneous change in the interparticle interaction. Excess amount of silica is added via mechanical grinding which results in a series of samples with volume fraction ($\varphi$) as 20\%, 15\%, 10\%, 7.5\% 5\%, 1\% and 0.1\%. This provides the platform to study the effect of variation of interparticle interactions on magnetic properties. From now on, different samples of Co/CoO embedded in SiO$_2$ matrix with a variation of interparticle interactions will be designated as CS-20, CS-15, CS-10, CS-7.5, CS-5, CS-1 and CS-0.1. For comparison, pure Co/CoO core/shell nanoparticle is named as CS-100. Fig. 1 demonstrate a schematic diagram, of changing volume fraction.

The structural characterization of the sample has been performed via Powder X-Ray diffraction (PXRD) pattern, recorded in a Bruker D8 Advanced Diffractometer using Cu K$\alpha$ ($\lambda$=1.54184 {\AA}) radiation source with a scan speed of 0.02$^{\circ}$/4 s. Actual shape, grain size and morphology of the NPs are assessed by Transmission Electron Microscope (TEM), equipped with an energy dispersive X-ray spectrometer (JEOL TEM, 2010). Temperature dependent dc magnetization measurements are performed via commercial SQUID magnetometer (MPMS-3). In the zero-field cooled (ZFC) protocol the sample is cooled in zero field and the magnetization is recorded in a static magnetic field during the heating cycle. In the field-cooled (FC) protocol sample is cooled in presence of a static magnetic field and magnetization measurements are performed either in heating or in cooling mode.

\begin{figure}[thpb]
\vskip 0.0 cm
\centering
\includegraphics[width=0.5\columnwidth]{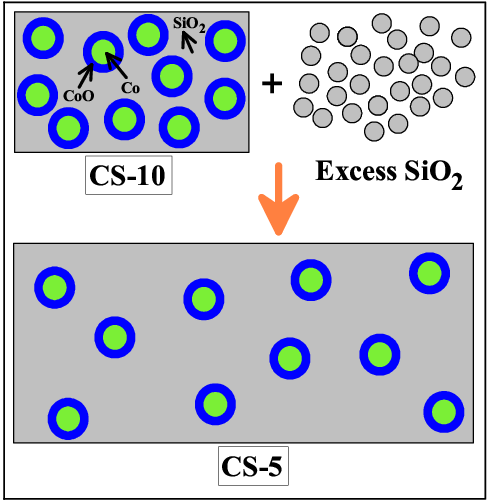}
\caption{\label{Fig.1} Schematic diagram representing change in volume fraction of Co-CoO nanoparticles from 10\% to 5\% by introducing excess silica matrix in the system.}

\end{figure}
\section{Results and discussions}

\subsection{Structural characterization}
\begin{figure}[thpb]
\vskip 0.0 cm
\centering
\includegraphics[width=0.5\columnwidth]{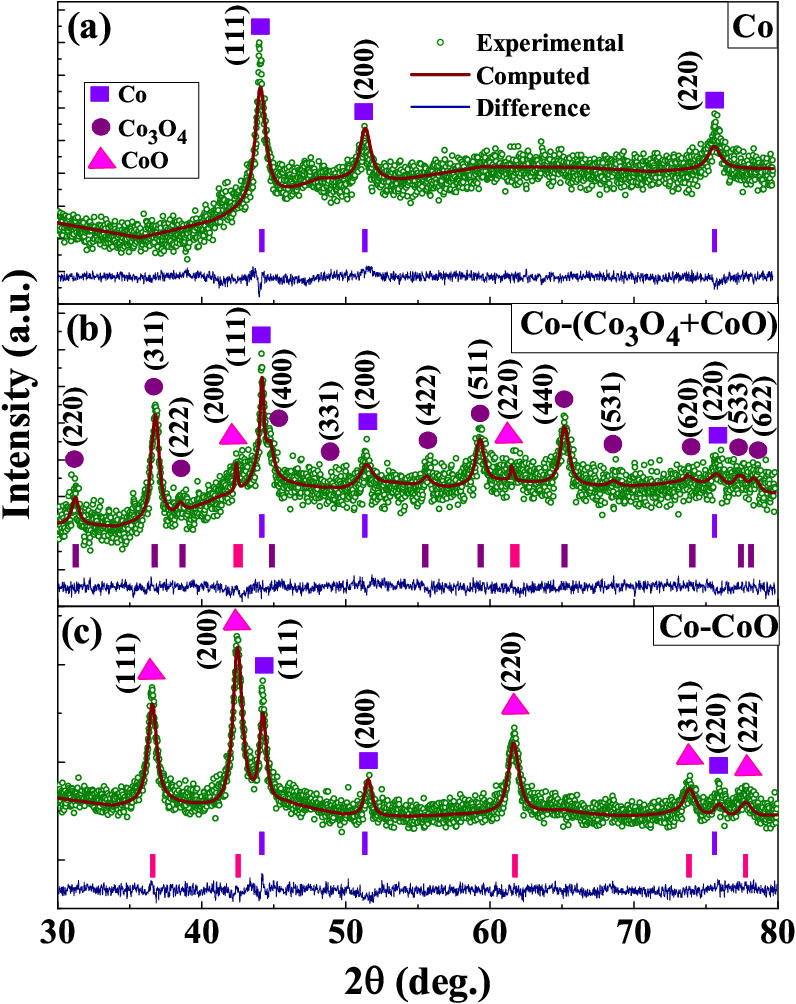}
\caption{\label{Fig.2} PXRD pattern of (a) Co nanoparticle, (b) Co-(Co$_3$O$_4$+CoO), after heating the Co nanoparticle in open air and (c) Co/CoO core/shell structure. Solid continuous curves are the fits using Rietveld refinement and the lowermost plots in each panel
are the residuals. Vertical bars correspond to different peak positions corresponding to different crystallites.}
\end{figure}

PXRD pattern of as-synthesized primary sample Co, intermediate sample Co-(Co$_3$O$_4$+CoO) and final product Co/CoO are recorded in the range of 30$^{\circ}$-80$^{\circ}$ at 300 K and are depicted in Fig.~\ref{Fig.2}(a)-(c). An elaborate Rietveld refinement has been performed on the diffraction patterns using MAUD (materials analysis using diffraction) software, considering face-centered $Fm3m$ space group for Co, CoO and $Fd3m$ for Co$_3$O$_4$. A close match of experimental data and computed curves are noticed as indicated by the difference plots at the bottom of Figs.~\ref{Fig.2}(a)-(c). Fig.~\ref{Fig.2} 2(a) corresponds to Co nanoparticle only whereas Fig.~\ref{Fig.2}(b) represents the XRD pattern of as synthesized Co-(Co$_3$O$_4$+CoO) developed after oxidation treatment Co nanoparticles in open air. Co-(Co$_3$O$_4$+CoO) demonstrate the characteristic peaks of Co, Co$_3$O$_4$ and CoO, which are in accordance with the JCPDS data (15-0806), (43-1003) and (43-1004), respectively. Weight percentage of different compositions present in this intermediate sample is Co:Co$_3$O$_4$:CoO$\approx$26:68:6, as evident from the refinement results. Fig.~\ref{Fig.2}(c) shows the peak positions corresponding to Co and CoO phases only. Proper indexing of all the peaks in Fig.~\ref{Fig.2}(c) rules out the possibility of presence of any other secondary phases or impurity in the final product Co/CoO. Refinement also suggests that the weight percentage of Co:CoO$\approx$20:80. This suggests, controlled oxidation-reduction converts Co$_3$O$_4$ into stable CoO phase \cite{Zhang}. Vertical bars at the bottom of Figs.~\ref{Fig.2}(a)-(c) in three different colors correspond to different phases (Co, Co$_3$O$_4$ and CoO). Information extracted from the Rietveld refinement such as, lattice parameters, atomic positions, refined parameters ($R_p$, $R_{wp}$ and $\chi^2$ as shown in Table 1), bond angles are in acceptable range and are in close agreement with the recent results \cite{Cantera,DeAlba,DeToro}. Increase in average crystallite sizes ($D$) of the three sample, as determined by modified Scherrer’s formula \cite{williamson,De2015} from the corresponding PXRD patterns, commensurate with the heat treatment (see Table 1). 

\begin{figure*}[t]
\vskip 0.0 cm
\centering
\includegraphics[width=0.95\columnwidth]{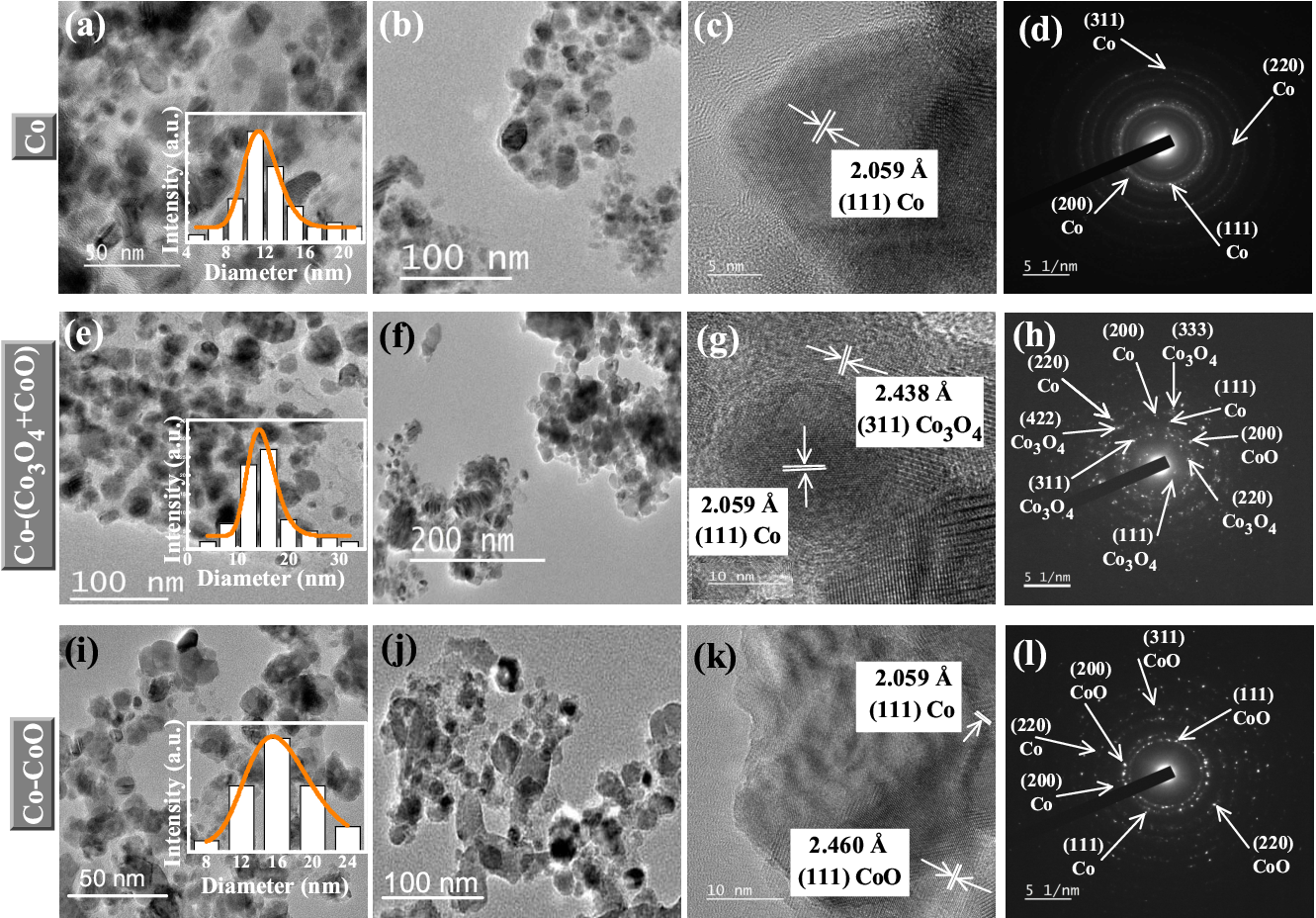}
\caption{\label{Fig.3} TEM images of Co ((a)\&(b)), Co-(Co$_3$O$_4$+CoO) ((e)\&(f)) and Co/CoO ((i)\&(j)) show well dispersed particles with distribution of particle sizes. Insets of (a), (e) and (i) depict the histograms of particle sizes, fitted with log-normal distribution function. (c), (g) and (k) HR TEM images, highlighting plane spacings of Co, Co$_3$O$_4$ and CoO. (d), (h) and (i) correspond to SAED pattern of the three samples, indicating different crystalline planes which are in accordance with the PXRD pattern.}
\end{figure*}

\begin{table*}[t]
\centering
\caption{Lattice and refinement parameters}
	\begin{adjustwidth}{-\extralength}{0cm}
		\newcolumntype{C}{>{\centering\arraybackslash}X}
		\begin{tabularx}{\fulllength}{CCCCCCCC}

\hline
\hline 
 Sample        &  Structure   &  a       & $R_p$ & $R_{wp}$ & $\chi^2$ & $D$ & $D$  \\
               &              &  ({\AA}) & (\%)  &  (\%)    &          & (PXRD)(nm)&  (TEM)(nm)      \\
\hline
Co             &  Fm3m          &  3.558(0.004)     &  1.8399   &   2.3351  &  1.1762  & 26  & 12.22(0.09)   \\ 
\hline
               & Fm3m (Co)        &   3.551(0.002)    &      &     &  &   & \\
Co/(Co$_3$O$_4$+CoO)    & Fd3m(Co$_3$O$_4$)&   8.093(0.002)    &  1.6499  &  2.0905   & 1.0839  & 29 &   16.57(0.27)           \\ 
							 & Fm3m (CoO )      &   4.260(0.001)    &    &             &          &   &               \\
\hline
Co/CoO            & Fm3m (Co) &   3.545(0.001)    &   1.3173   &  1.6411   &  1.1198  & 30 &  17.54(0.02) \\
               & Fm3m (CoO)&   4.255(0.001)    &            &               &          &         &     \\
\hline      
\hline         
\end{tabularx}
\end{adjustwidth}
\end{table*}
To investigate the actual size and morphology of the samples and its procedural changes with synthesis, TEM micrographs are studied for all the three samples. Figs~\ref{Fig.3}(a-b), (e-f) and (i-j) depict spherical nanoparticles for Co, Co-(Co$_3$O$_4$+CoO) and Co/CoO, respectively. Insets of Figs. \ref{Fig.3}(a), (e) and (i) represent the histograms of particle size distributions of the samples, fitted with log normal distribution function. Fittings result average particle sizes that are found to be $\approx$12 nm for Co ($\sigma_{log}$=0.15), $\approx$15 nm for Co-(Co$_3$O$_4$+CoO) ($\sigma_{log}$=0.18) and $\approx$16 nm for Co/CoO ($\sigma_{log}$=0.23), respectively. Here too, the gradual increase in particle sizes supports the findings of the PXRD studies and is in accordance with the heat treatment. Figs.~\ref{Fig.3}(c), (g) and (k) represent the high resolution (HR) TEM images of the samples revealing formation of high crystallinity up to the edges of the particles. Lattice plane spacing of (111) plane for Co, (311) for Co$_3$O$_4$ and (111) for CoO are observed in different HR-TEM images which corresponds to the formation of core/shell structure. Selected area electron diffraction (SAED) patterns of the nanoparticles are presented in Figs.~\ref{Fig.3}(d), (h) and (l) where planes corresponding to Co, Co$_3$O$_4$ and CoO are observed. Findings support the PXRD studies and no impurity plane is noted. The results obtained from two different techniques replicate almost similar results, indicating the purity of the samples. Then, the duly characterized core/shell Co/CoO nanoparticles undergo a treatment of variation of interparticle interactions via introduction of SiO$_2$ in desired limit and samples with different volume fractions are prepared for detailed magnetic characterization.

\subsection{Magnetic characterization}
\subsubsection{ZFC-FC thermal dependence}
\begin{figure}[htpb]
\vskip 0.0 cm
\centering
\includegraphics[width=0.5\columnwidth]{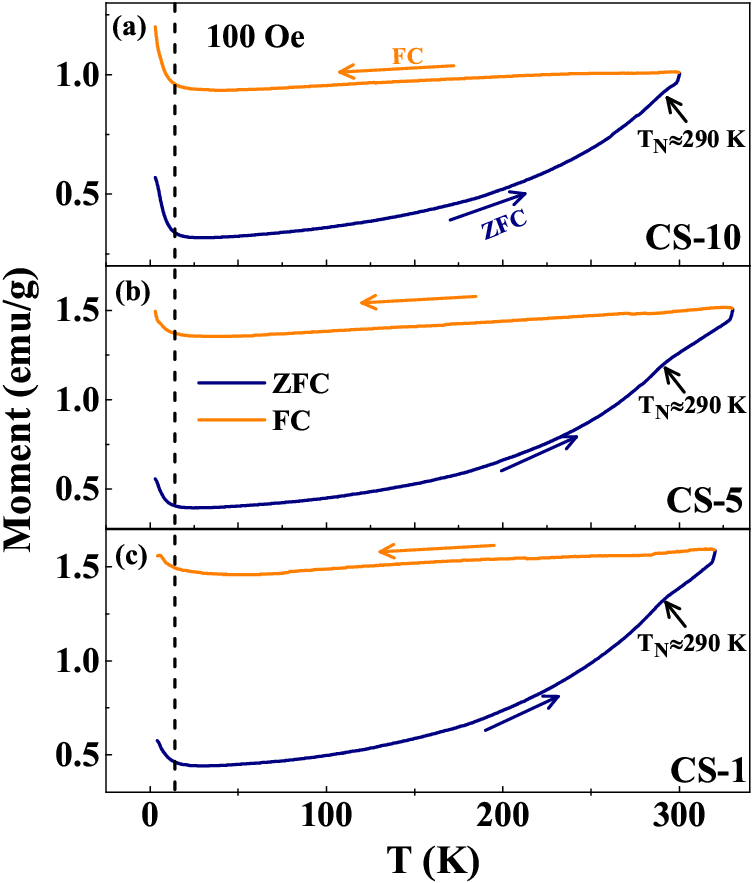}
\caption{\label{Fig.4} Thermal dependence of the ZFC and FC magnetization of samples (a) CS-10, (b) CS-5 and (c) CS-1, at $100$ Oe. Anomalies evident near $290$ K  correspond to the Néel temperature ($T_N$) of CoO. In the low temperature region, a Cuire-tail like behavior is observed for all the samples.}
\end{figure}

To understand the effect of interparticle interactions on the magnetic behavior of as synthesized Co/CoO nanoparticles with different volume fractions, thermal variation of magnetization ($M-T$) in ZFC and FC modes in an applied field of 100 Oe field were measured and the results are shown Fig.~\ref{Fig.4} for CS-10, CS-5 and CS-1 samples as representative of the series. 
The general trends of $M-T$ curves is similar for the three samples, showing irreversibility up to the maximum measured temperature of $320$ K \cite{DEJAP} and suggesting that the blocking temperature  is above this value which is reasonable given the relatively big nanoparticle sizes. All the curves decrease monotonously below $320$ K, but ZFC curves show a subtle but noticeable anomaly at $\approx 290$ K, which corresponds to the Neél temperature ($T_N$) of the AFM CoO shell \cite{Simeonidis,Feygenson,Tracy} and explains the steeper decrease of the magnetization below this temperature. No anomaly is observed at the ordering temperature of Co$_3$O$_4$ ($\approx 40$K), confirming the absence of this phase as also evidenced by PXRD.
We notice a sudden magnetization increase below $12$ K for all the samples, generally known as Curie-tail like behavior \cite{Chattopadhyay1}, that is usually observed in systems with broken spin chains or paramagnetic like impurities \cite{Chattopadhyay}.  Recent reports suggest that the presence of oxygen atoms in the CoO shell generate holes that break the infinite lattice chain and thus can lead to the low temperature uptail \cite{Sirker}. 
The Curie-tail is more pronounced as the volume fraction increases, indicating that upon dilution collective magnetic effects diminish leading to a reduction of the Curie-tail. 
Curves of samples with increasing $\varphi$ in Fig.~\ref{Fig.4} display a progressive decrease of the magnitude of the magnetization that is a first evidence that interparticle interactions change the magnetic behavior of the core/shell nanoparticles.

\subsubsection{ZFC Hysteresis loops}
\begin{figure}[t]
\vskip 0.0 cm
\centering
\includegraphics[width=0.5\columnwidth]{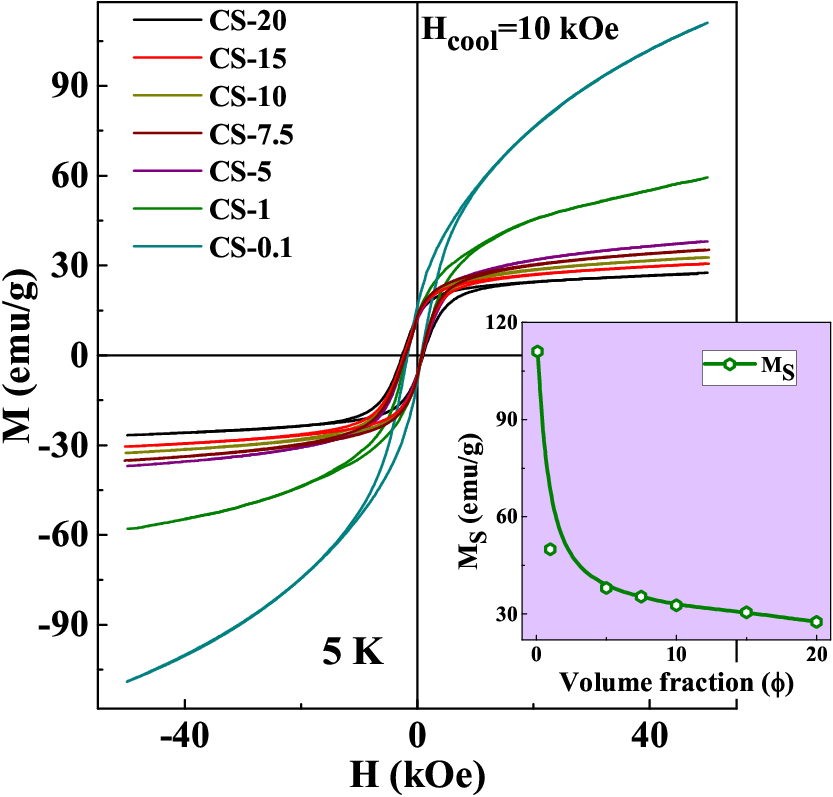}
\caption{\label{Fig.6} $M-H$ loops of samples CS-20, CS-15, CS-10, CS-7.5 CS-5, CS-1, CS-0.1 measured in ZFC mode. Inset shows the variation of $M_S$ with volume fraction ($\varphi$).}
\end{figure}

In order to study the influence of interactions on the reversal by a magnetic field, $M-H$ hysteresis loops for all the seven samples with different $\varphi$ have been measured in between $\pm$50 kOe magnetic fields after cooling the samples from 300 K to 4 K in zero field (ZFC), as displayed in
Fig.~\ref{Fig.6}. 
For the most diluted samples (CS-0.1, CS-1), the loops do not saturate even at $50$ kOe. The observed high field quasi linear response characteristic of a SPM is a typical signature of a disordered system. In these cases, the response to a magnetic field as reflected in the loops must come from intrinsic properties of an individual core/shell particles. Surface spins may present frustration due to broken links or lack of coordination, and increased surface anisotropy \cite{IglesiasKachkachi}, which distinguish them from those at inner regions of the shell that are pinned by the coupling to core spins and do not contribute to the magnetic response \cite{Simeonidis}.
Therefore, the absence of saturation at low temperature with lower $\varphi$ can be attributed to the progressive alignment alignment of the outermost layer of the AFM shell spins towards the core magnetization.

However, this behavior changes to $M-H$ loops that saturate towards lower values of the magnetization as the interparticle separation is decreased. This can be seen in Fig. \ref{Fig.6} for samples CS-5 to CS-20, where in the latter the magnetization saturates already at $10$ kOe to $M_S\approx 27$ emu/g, whereas for the most diluted system CS-0.1 $M_S\approx$111 emu/g at 50 kOe field. As $\varphi$ increases, the mean interparticle separation decreases and dipolar interactions become more relevant. Thus, in the more concentrated assemblies, the magnetic behavior is dominated by collective effects induced by dipolar interactions of the FM Co cores, the $M-H$ loops display the typical shape of a superferromagnetic (SFM) system \cite{Bedanta} and the SPM contribution coming from the individual nanoparticle response is suppressed. 
Assuming that dipolar interactions arise from the FM cores (neglecting the AF shell contribution to the total magnetization),  typical dipolar energies between two Co particles separated by a distance $d$ twice the shell thickness can be evaluated as $E_{\text{dip}}=\frac{\mu_0}{4\pi}\frac{M_s^2V^2}{d^3}\approx 1220$ K, to be compared with anisotropy energy $E_{\text{ani}}=KV\approx 3030$ K, which qualifies our samples with higher volume fraction as governed by collective dipolar behavior \cite{Sanchez,Sanchez2}.
The overall behavior is consistent with the changes in the magnetization in the low temperature region of the $M-T$ curves shown in Fig.~\ref{Fig.4}.

\subsubsection{Exchange bias}
\begin{figure*}[tb]
\vskip 0.0 cm
\centering
\includegraphics[width=0.9\columnwidth]{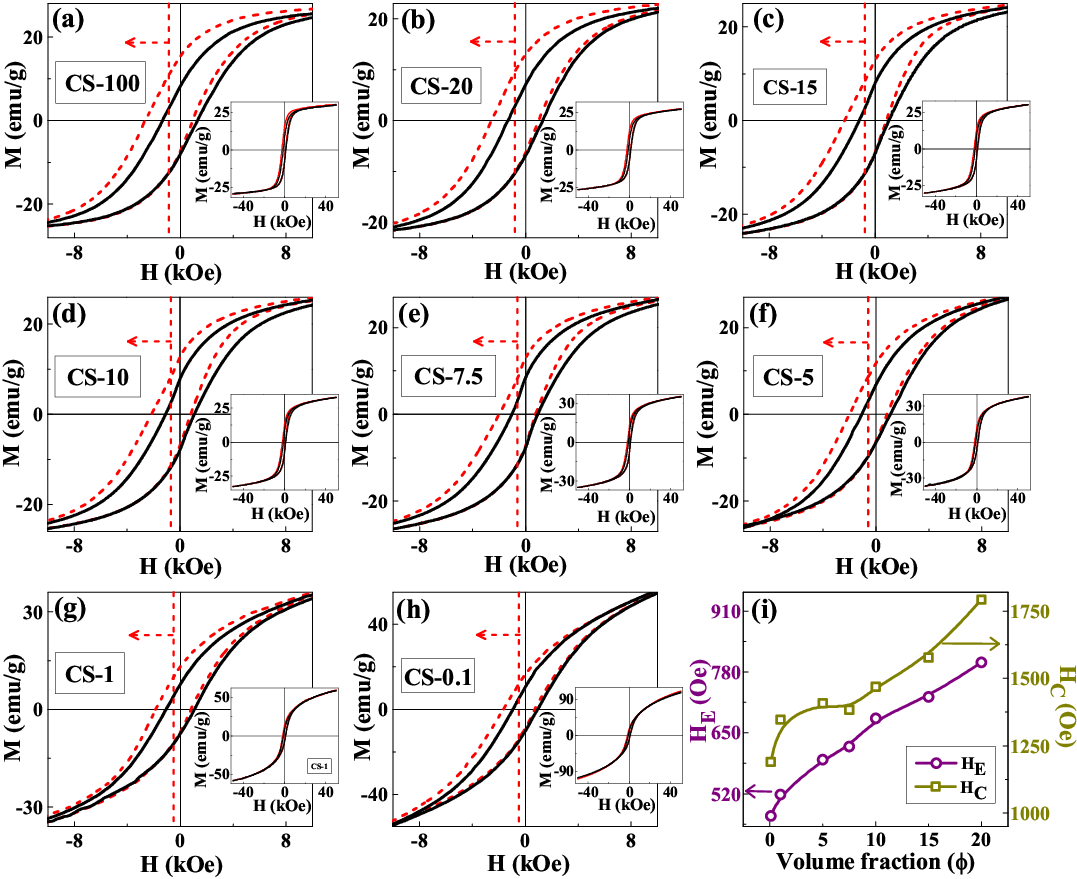}
\caption{\label{Fig.5} (a)-(e) Central portions of the ZFC (solid line) and in the FC ($H_{cool}$=10 kOe, dashed line) $M-H$ loops recorded at 4 K for CS-100, CS-20, CS-15, CS-10, CS-7.5 CS-5, CS-1 and CS-0.1. Full scale $M-H$ loops are depicted in the corresponding insets. (i) Variation of coercivity ($H_C$) and EB field ($H_E$) with change in volume fraction ($\varphi$).}
\end{figure*}

The influence of interactions on the EB effect was evidenced by recording also hysteresis loops after cooling in an applied field $H_{cool}$=10 kOe. The central portions of the loops are shown by red dashed lines in Figs.~\ref{Fig.5}(a-h); ZFC loops are displayed in the same figure as black solid lines. For comparison, the $M-H$ loop of CS-100, (Co/CoO without SiO$_2$) is also included and insets in the corresponding panels show the $M-H$ loops in full scale.
All loops after FC show a notable shift contrary to the cooling field direction as indicated by the arrows with a concomitant increase in the coercive field with respect to the one measured under ZFC conditions \cite{Iglesias}. The fact that the shift is observed even for the most diluted sample (CS-0.1) asserts the existence of a unidirectional anisotropy, originated by the freezing of the AF shell spins, and that the resulting EB effect is induced at the individual particle level \cite{Giri2011,Nogues}. 
The EB field ($H_E$) and coercivity ($H_C$) have been determined as $H_E$=$\left|H_2+H_1\right|$/2 and $H_C$=$\left|(H_2-H_1)\right|$/2; where $H_1$ and $H_2$ are the coercivities of the decreasing and decreasing field branches respectively \cite{De2016}.
The resulting dependence of $H_E$ and $H_C$ with the volume fraction $\varphi$ in the different samples is depicted in Fig.~\ref{Fig.5}(i), which shows a monotonic increase of both  quantities with increasing $\varphi$, i.e. increasing interparticle interactions.
In going from CS-0.1 to CS-20 $H_C$ and $H_E$ change from $1191$ Oe to $1793$ and from $473$ to $800$, respectively. 
Since the samples with different $\varphi$ are derived from the same mother sample (CS-100) by introduction of additional SiO$_2$, it can be assumed that the contribution to the loop shift coming from the exchange coupling at the interfacial region of the individual particles is the same for all of them. Therefore, the reason for the observed notably higher values $H_C$, $H_E$ at higher concentrations must have its origin in the decrease of interparticle distance at higher $\varphi$.
On the one hand, it can be argued that when CoO shells come close to each other, the effective thickness of AFM shell increases, leading to an increase in magnetic coupling and coercive field between the FM core and AFM shell as argued in \cite{Nogues2006}. 
However, this effect could only partially explain the observations. Recent MC simulations of a simplified macrospin model of core/shell nanoparticle assemblies \cite{Kostopoulou2014,Silva2022} have shown that the EB field is influenced by both direct interparticle exchange and dipolar interactions, whose contributions could be separately evaluated. Simulation results were in good agreement with experimental results showing an increase of $H_{E}$ in powder samples compared to diluted ferrofluids \cite{Silva2022,Omelyanchik2021}.

We believe that, due to the higher core sizes and shell thicknesses of our samples, the reason behind the EB enhancement is the increase of the dipolar fields felt by the individual particles as the particles approach to each other. 
In an individual core/shell nanoparticle, the loop shift is related to the local exchange field created by the uncompensated spins at the interface \cite{Iglesias1} that adds in opposite directions at the decreasing and increasing field loop branches and generates a unidirectional anisotropy. 
Our situation bears similarities with the dipole-induced EB model proposed in \cite{Torres2017}  to explain EB in AFM/FM thin films separated by an interface layer.
When the particle is in an assembly, the dipolar field generated by the rest of particles superposes to the local exchange field and acts as an additional source of unidirectional anisotropy that enhances the EB effect.
In order to reinforce this interpretation, we will now give an estimation of typical dipolar fields that can be found in SFM samples, using an argument that was also employed to explain the shift in energy barrier distributions due to dipolar interactions in \cite{Moya}. 
Let us consider a Co core of diameter $D= 18$ nm. A rough estimate of the dipolar field felt at a distance two times the typical shell thickness $d= 24$ nm can be obtained as $H_{\text{dip}}=\frac{\mu_0}{4\pi}\frac{M_sV}{d^3}\approx 110$ Oe. This is of the correct order of magnitude of the increase in $327$ Oe observed for $\varphi= 20$ in Fig.\ref{Fig.5}(i), if we consider that the $H_{\text{dip}}$ acting on a particle receives contributions from several neighbors and that their magnetizations may not be aligned. 

\begin{figure}[bhtp]
\vskip 0.0 cm
\centering
\includegraphics[width=0.75\columnwidth]{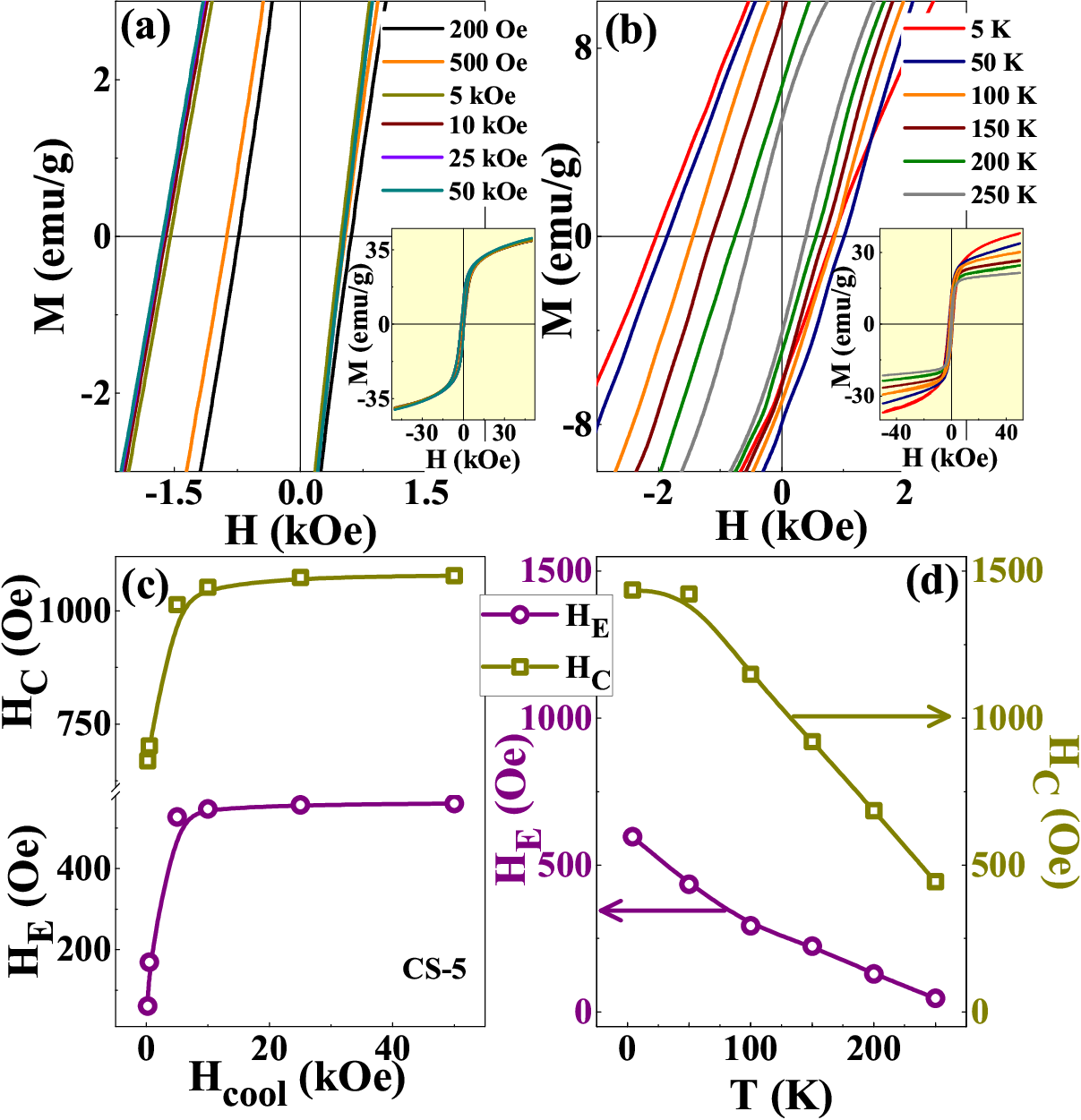}
\caption{\label{Fig.7} (a) Low field region of $M-H$ loops of CS-5 recorded at 4 K in between $\pm$50 kOe for $H_{\text{cool}}$=0.2, 0.5, 5, 10, 25 and 50 kOe. Inset of (a) shows the corresponding $M-H$ loops in full scale. (b) $M-H$ loops of CS-5 measured at different temperatures in FC mode at $H_{cool}$=10 kOe. Inset of (b) shows the corresponding $M-H$ loops in full scale. (c) and (d) Variation of $H_E$ and $H_C$ with change in cooling field and temperature, respectively.}
\end{figure}

To study the dependence of $H_C$ and $H_E$ with cooling field ($H_{\text{cool}}$), $M-H$ loops at different fields ($H_{\text{cool}}$=0.2, 0.5, 5, 10, 25, 50 kOe) were recorded in between $\pm$50 kOe after cooling the sample from $300$ to $4$ K in FC mode. The results are shown in Fig. \ref{Fig.7}(a) for the CS-5 sample as representative of the series. Hysteresis loops at $H_{\text{cool}}=10$ kOe were also recorded for a wide temperature range (5-250 K) to study the nature of EB for CS-5 at different temperatures (refer Fig.~\ref{Fig.7}(b)). 
The variation of the calculated parameters $H_E$ and $H_C$ with $H_{\text{cool}}$ and $T$ are presented in Figs.~\ref{Fig.7}(c) and (d), respectively. 
Both $H_C$ and $H_E$ increase with $H_{\text{cool}}$ and almost become saturated near $10$ kOe. Then, we can exclude that observed phenomenology can be attributed to a minor loop effect, in accordance with recent reports \cite{De2016,Goswami-JALCOM}. 
Fig.~\ref{Fig.7}(d) demonstrates a monotonic decrease of $H_C$ and $H_E$ with increase in temperature. $H_E$ almost vanishes at $250$ K which is close to the N$\acute{e}$el temperature of CoO shell. Decrease of $H_E$ with $T$ may be due to the loss of interface coupling between Co core and CoO shell caused by the increase in thermal fluctuations and the decrease in AFM anisotropy with increase in temperature \cite{Eftaxias,Sort}. 

\section{Conclusions}
 
Co nanoparticle has been synthesized using conventional sol-gel technique with average particle size $12$ nm. CoO shell has been formed over the Co nanoparticle via controlled oxidation-reduction which results in the formation of core/shell nanoparticles (Co/CoO) with average particle sizes $18$ nm. Samples were characterized by PXRD and TEM analysis. Coexistence of Co and CoO phases are confirmed without a very regular core/shell structure. Interparticle interaction/separation among the individual nanoparticles has been tuned by changing the volume fraction via introduction of additional SiO$_2$ matrix. Notably, different trends in the thermal dependence of the magnetization are observed between samples with different interparticle interactions. 
Starting from the most concentrated sample, a gradual decrease in coercivity and increase in non-saturation tendency was obtained with the incorporation of the non-magnetic matrix from the $M-H$ loops, demonstrating the change from SPM response due to individual nanoparticles to a SFM collective behavior as the interparticle interactions increased. 
Our study of EB effects, have revealed that $H_C$, $H_E$ and $M_S$ can be tuned monotonously with the increase of the volume fraction of the core/shell nanoparticles.
We have given an interpretation of the nature of the variations of $H_C$ and $H_E$ with $\varphi$, linking them to changes in interfacial coupling of FM core and AFM shell and increase of the local fields felt by the Co cores as a consequence of increasing dipolar interactions when the distance between the nanoparticles is decreased. 

It is worth noticing that the understanding of the variation of EB related parameters with interparticle interactions along with our previous findings of variation of EB with core to shell diameter ratio \cite{De2016} provide us a platform to tune or have a deep control over the EB related phenomenon and parameters of core/shell structures to explore application oriented device fabrication. 
In this study, Co/CoO nanostructure has been chosen as a representative of metal/metal oxide (FM/AFM) system revealing EB effect. The dependency of $H_C$ and $H_E$ reported here with interparticle interaction, may be considered as a general phenomenon after similar studies with core/shell nanoparticles of similar and different compositions.
Finally, we may conclude that, though the EB phenomenon was discovered about $70$ years ago, there are still many out of the box origins of EB mechanism that differ from the conventional knowledge of pinning mechanism at the interface between materials with different magnetic ordering. Full understanding of these new underlying mechanisms  in EB assemblies will necessitate further experimental studies and will need to improve current theoretical frameworks that can incorporate collective effects due to dipolar interactions.

\authorcontributions{ Conceptualization, D.D.; methodology, S.G., P.G., S.N., S.B., M.C. and D.D.;
validation, S.B., Ò.I. and D.D.; formal analysis, S.G., P.G., S.N., S.B., Ò.I., M.C. and D.D.; investigation,
S.G., P.G., S.N., S.B., M.C. and D.D.; data curation, S.G., P.G. and S.N.; writing—original draft
preparation, S.G., P.G., S.N., S.B., Ò.I., M.C. and D.D.; writing—review and editing, S.B., Ò.I., M.C.
and D.D.; visualization, S.B., D.D. and Ò.I.; supervision, S.B., M.C. and D.D.; project administration,
D.D.; funding acquisition, S.B., Ò.I., M.C. and D.D. All authors have read and agreed to the published
version of the manuscript.}

\funding{This work is an outcome of collaborative research of Material Science Research Lab, The Neotia University and Laboratory for Nanomagnetism and Magnetic Materials, NISER, Bhubaneswar. D. De, M. Chakraborty and S. Goswami thank SERB Project EMR/2017/001195 for financial support. S. Bedanta wishes to thank DAE for financial support. Ò. Iglesias thanks Spanish MINECO projects PGC2018-097789-B-I00, PID2019-109514RJ-I00 and the European Union FEDER funds.}

\acknowledgments{D. De, M. Chakraborty and S. Goswami thank The Neotia University authority \& Sukumar Sengupta Mahavidyalaya for their cooperation and encouragement. All authors thank, Prof. N. Sarkar, The Neotia University, Dr. S. Dey, Purulia Polytechnic, W.B., India, Mr. A. Maity of The Neotia University and Dr. A. Banerjee, IACS Kolkata for research support.}

\conflictsofinterest{The authors declare no conflict of interest.} 
\begin{adjustwidth}{-\extralength}{0cm}

\reftitle{References}


\begin{thebibliography}{999}

\bibitem[He \em{et~al.}(2013)He, Zhong, Au, and Du]{He}
He, X.; Zhong, W.; Au, C.T.; Du, Y.
\newblock Size dependence of the magnetic properties of Ni nanoparticles
  prepared by thermal decomposition method.
\newblock {\em Nanoscale Research Lett.} {\bf 2013}, {\em 8},~446.
\newblock {\url{https://doi.org/10.1186/1556-276X-8-446}}.

\bibitem[Goswami \em{et~al.}(2020)Goswami, Manna, Bedanta, Dey, Chakraborty,
  and De]{Goswami}
Goswami, S.; Manna, P.K.; Bedanta, S.; Dey, S.K.; Chakraborty, M.; De, D.
\newblock Surface driven exchange bias in nanocrystalline CoCr$_2$O$_4$.
\newblock {\em J. Phys. D: Appl. Phys.} {\bf 2020}, {\em 53},~305303.
\newblock {\url{https://doi.org/10.1088/1361-6463/ab87c7}}.

\bibitem[Auvinen \em{et~al.}(2011)Auvinen, Alatalo, Haario, Jalava, and
  Lamminm$\ddot{a}$ki]{Auvinen}
Auvinen, S.; Alatalo, M.; Haario, H.; Jalava, J.P.; Lamminm$\ddot{a}$ki, R.J.
\newblock Size and Shape Dependence of the Electronic and Spectral Properties
  in TiO$_2$ Nanoparticles.
\newblock {\em J. Phys. Chem. C} {\bf 2011}, {\em 115},~8484.
\newblock {\url{https://doi.org/10.1021/jp112114p}}.

\bibitem[Ashraf \em{et~al.}(2018)Ashraf, Peng, Zare, and Rhee]{Ashraf}
Ashraf, M.A.; Peng, W.; Zare, Y.; Rhee, K.Y.
\newblock Effects of Size and Aggregation/ Agglomeration of Nanoparticles on
  the Interfacial/Interphase Properties and Tensile Strength of Polymer
  Nanocomposites.
\newblock {\em Nanoscale Research Lett.} {\bf 2018}, {\em 13},~214.
\newblock {\url{https://doi.org/10.1186/s11671-018-2624-0}}.

\bibitem[Papaefthymiou \em{et~al.}(2009)Papaefthymiou, Devlin, Simopoulos, Yi,
  Riduan, Lee, and Ying]{Papaefthymiou}
Papaefthymiou, G.C.; Devlin, E.; Simopoulos, A.; Yi, D.K.; Riduan, S.N.; Lee,
  S.S.; Ying, J.Y.
\newblock Interparticle interactions in magnetic core/shell nanoarchitectures.
\newblock {\em Phys. Rev. B} {\bf 2009}, {\em 80},~024406.
\newblock {\url{https://doi.org/10.1103/PhysRevB.80.024406}}.

\bibitem[De \em{et~al.}(2016)De, Iglesias, Majumdar, and Giri]{De2016}
De, D.; Iglesias, {\`{O}}.; Majumdar, S.; Giri, S.
\newblock Probing core and shell contributions to exchange bias in
  Co/Co$_3$O$_4$ nanoparticles of controlled size.
\newblock {\em Phys. Rev. B} {\bf 2016}, {\em 94},~184410.
\newblock {\url{https://doi.org/10.1103/PhysRevB.94.184410}}.

\bibitem[Chaudhuri and Paria(2012)]{chaudhuri}
Chaudhuri, R.G.; Paria, S.
\newblock Core/Shell Nanoparticles: Classes, Properties, Synthesis Mechanisms,
  Characterization, and Applications.
\newblock {\em Chem. Rev.} {\bf 2012}, {\em 112},~2373.
\newblock {\url{https://doi.org/10.1021/cr100449n}}.

\bibitem[Sch$\ddot{a}$rtl(2010)]{Schartl}
Sch$\ddot{a}$rtl, W.
\newblock Current directions in core–shell nanoparticle design.
\newblock {\em Nanoscale} {\bf 2010}, {\em 2},~829.
\newblock {\url{https://doi.org/10.1039/c0nr00028k}}.

\bibitem[Al-Ogaidi \em{et~al.}(2014)Al-Ogaidi, Gou, Al-kazaz, Aguilar,
  Melconian, Zheng, and Wu]{Al-Ogaidi}
Al-Ogaidi, I.; Gou, H.; Al-kazaz, A.K.A.; Aguilar, Z.P.; Melconian, A.K.;
  Zheng, P.; Wu, N.
\newblock A gold\@silica core–shell nanoparticle-based surface-enhanced Raman
  scattering biosensor for label-free glucose detection.
\newblock {\em Anal. Chim. Acta} {\bf 2014}, {\em 811},~76.
\newblock {\url{https://doi.org/10.1016/j.aca.2013.12.009}}.

\bibitem[Cha \em{et~al.}(2015)Cha, Mun, Chang, Kim, Kim, Jin, Lee, Shin, Kim,
  and Kim]{Cha}
Cha, S.K.; Mun, J.H.; Chang, T.; Kim, S.Y.; Kim, J.Y.; Jin, H.M.; Lee, J.Y.;
  Shin, J.; Kim, K.H.; Kim, S.O.
\newblock Au-Ag Core-Shell Nanoparticle Array by Block Copolymer Lithography
  for Synergistic Broadband Plasmonic Properties.
\newblock {\em ACS Nano} {\bf 2015}, {\em 9},~5536.
\newblock {\url{https://doi.org/10.1021/acsnano.5b01641}}.

\bibitem[Feygenson \em{et~al.}(2010)Feygenson, Yiu, Kou, Kim, and
  Aronson]{Feygenson}
Feygenson, M.; Yiu, Y.; Kou, A.; Kim, K.S.; Aronson, M.C.
\newblock Controlling the exchange bias field in Co core/CoO shell
  nanoparticles.
\newblock {\em Phys. Rev. B} {\bf 2010}, {\em 81},~195445.
\newblock {\url{https://doi.org/10.1103/PhysRevB.81.195445}}.

\bibitem[Meiklejohn and Bean(1956)]{Meiklejhon}
Meiklejohn, W.H.; Bean, C.P.
\newblock New Magnetic Anisotropy.
\newblock {\em Phys. Rev.} {\bf 1956}, {\em 102},~1413.
\newblock {\url{https://doi.org/10.1103/PhysRev.102.1413}}.

\bibitem[Vasilakaki and Trohidou(2009)]{Vasilakaki}
Vasilakaki, M.; Trohidou, K.N.
\newblock Numerical study of the exchange-bias effect in nanoparticles with
  ferromagnetic core/ ferrimagnetic disordered shell morphology.
\newblock {\em Phys. Rev. B} {\bf 2009}, {\em 79},~144402.
\newblock {\url{https://doi.org/10.1103/PhysRevB.79.144402}}.

\bibitem[Salazar-Alvarez \em{et~al.}(2007)Salazar-Alvarez, Sort,
  Suri$\tilde{n}$ach, Bar$\acute{o}$, and Nogu$\acute{e}$s]{Alvarez}
Salazar-Alvarez, G.; Sort, J.; Suri$\tilde{n}$ach, S.; Bar$\acute{o}$, M.D.;
  Nogu$\acute{e}$s, J.
\newblock Synthesis and Size-Dependent Exchange Bias in Inverted Core-Shell
  MnO|Mn$_3$O$_4$ Nanoparticles.
\newblock {\em J. Am. Chem. Soc.} {\bf 2007}, {\em 129},~9102.
\newblock {\url{https://doi.org/10.1021/ja0714282}}.

\bibitem[Sahoo \em{et~al.}(2016)Sahoo, Giri, Dasgupta, Poddar, and Nath]{Sahoo}
Sahoo, R.C.; Giri, S.K.; Dasgupta, P.; Poddar, A.; Nath, T.K.
\newblock Exchange bias effect in ferromagnetic LaSrCoMnO$_6$ double
  perovskite: consequence of spin glass-like ordering at low temperature.
\newblock {\em J. Alloys and Compounds} {\bf 2016}, {\em 658},~1003.
\newblock {\url{https://doi.org/10.1016/j.jallcom.2015.11.025}}.

\bibitem[Wang \em{et~al.}(2004)Wang, Zhu, Zhao, Wang, Wang, Wang, and
  Zhan]{Wang}
Wang, H.; Zhu, T.; Zhao, K.; Wang, W.N.; Wang, C.S.; Wang, Y.J.; Zhan, W.S.
\newblock Surface spin glass and exchange bias in Fe$_3$O$_4$ nanoparticles
  compacted under high pressure.
\newblock {\em Phys. Rev. B} {\bf 2004}, {\em 70},~092409.
\newblock {\url{https://doi.org/10.1103/PhysRevB.70.092409}}.

\bibitem[Giri \em{et~al.}(2016)Giri, Sahoo, Dasgupta, Poddar, and Nath]{SKGiri}
Giri, S.K.; Sahoo, R.C.; Dasgupta, P.; Poddar, A.; Nath, T.K.
\newblock Giant spontaneous exchange bias effect in
  Sm$_{1.5}$Ca$_{0.5}$CoMnO$_6$ perovskite.
\newblock {\em J. Phys. D: Appl. Phys.} {\bf 2016}, {\em 49},~165002.
\newblock {\url{https://doi.org/10.1088/0022-3727/49/16/165002}}.

\bibitem[Giri \em{et~al.}(2011)Giri, Patra, and Majumdar]{Giri2011}
Giri, S.; Patra, M.; Majumdar, S.
\newblock Exchange bias effect in alloys and compounds.
\newblock {\em J. Phys.: Condens. Matter} {\bf 2011}, {\em 23},~073201.
\newblock {\url{https://doi.org/10.1088/0953-8984/23/7/073201}}.

\bibitem[Nogu\'es \em{et~al.}(2005)Nogu\'es, Sort, Langlais, Skumryev,
  Suri$\tilde{n}$ach, Mu$\tilde{n}$oz, and Bar$\acute{o}$]{Nogues}
Nogu\'es, J.; Sort, J.; Langlais, V.; Skumryev, V.; Suri$\tilde{n}$ach, S.;
  Mu$\tilde{n}$oz, J.S.; Bar$\acute{o}$, M.D.
\newblock Exchange bias in nanostructures.
\newblock {\em Phys. Rep.} {\bf 2005}, {\em 422},~65.
\newblock {\url{https://doi.org/10.1016/j.physrep.2005.08.004}}.

\bibitem[Goswami \em{et~al.}(2022)Goswami, Gupta, Bedanta, Chakraborty, and
  De]{Goswami-JALCOM}
Goswami, S.; Gupta, P.; Bedanta, S.; Chakraborty, M.; De, D.
\newblock Coexistence of exchange bias and memory effect in nanocrystalline
  CoCr$_2$O$_4$.
\newblock {\em J. Alloys and Compounds} {\bf 2022}, {\em 890},~161916.
\newblock {\url{https://doi.org/10.1016/j.jallcom.2021.161916}}.

\bibitem[Nayak \em{et~al.}(2020)Nayak, Manna, Vijayabaskaran, Singh, Chelvane,
  and Bedanta]{Bedanta3}
Nayak, S.; Manna, P.K.; Vijayabaskaran, T.; Singh, B.B.; Chelvane, J.A.;
  Bedanta, S.
\newblock Exchange bias in Fe/ Ir$_{20}$ Mn$_{80}$ bilayers: Role of spin-glass
  like interface and bulk’ antiferromagnet spins.
\newblock {\em J. Magn. Magn. Mater.} {\bf 2020}, {\em 499},~166267.
\newblock {\url{https://doi.org/10.1016/j.jmmm.2019.166267}}.

\bibitem[Nayak \em{et~al.}(2021)Nayak, Manna, Singh, and Bedanta]{Bedanta4}
Nayak, S.; Manna, P.K.; Singh, B.B.; Bedanta, S.
\newblock Effect of spin glass frustration on exchange bias in NiMn/CoFeB
  bilayers.
\newblock {\em Phys. Chem. Chem. Phys.} {\bf 2021}, {\em 23},~6481.
\newblock {\url{https://doi.org/10.1039/d0cp05726f}}.

\bibitem[Skumryev \em{et~al.}(2003)Skumryev, Stoyanov, Zhang, Hadjipanayis,
  Givord, and Nogu\'es]{Skumryev}
Skumryev, V.; Stoyanov, S.; Zhang, Y.; Hadjipanayis, G.; Givord, D.; Nogu\'es,
  J.
\newblock Beating the superparamagnetic limit with exchange bias.
\newblock {\em Nature} {\bf 2003}, {\em 423},~850.
\newblock {\url{https://doi.org/10.1038/nature0168}}.

\bibitem[Sharma \em{et~al.}(2016)Sharma, Albisetti, Monticelli, Bertacco, and
  Petti]{Sharma}
Sharma, P.P.; Albisetti, E.; Monticelli, M.; Bertacco, R.; Petti, D.
\newblock Exchange Bias Tuning for Magnetoresistive Sensors by Inclusion of
  Non-Magnetic Impurities.
\newblock {\em Sensor} {\bf 2016}, {\em 16},~1030.
\newblock {\url{https://doi.org/10.3390/s16071030}}.

\bibitem[Nogu\'es and Schuller(1999)]{Nogues1999}
Nogu\'es, J.; Schuller, I.K.
\newblock Exchange bias.
\newblock {\em J. Magn. Magn. Mater.} {\bf 1999}, {\em 192},~203.
\newblock {\url{https://doi.org/10.1016/S0304-8853(98)00266-2}}.

\bibitem[Akta{\c{s}} \em{et~al.}(2020)Akta{\c{s}}, Kocaman, and Basaran]{Aktas}
Akta{\c{s}}, K.Y.; Kocaman, B.; Basaran, A.C.
\newblock Magnetic and Electrical (GMR) Properties of Rh(IrMn)/Co/Cu/Ni(Py)
  Multilayered Thin Films.
\newblock {\em Journal of Superconductivity and Novel Magnetism} {\bf 2020},
  {\em 33},~2093.
\newblock {\url{https://doi.org/10.1007/s10948-020-05464-8}}.

\bibitem[Huang \em{et~al.}(2008)Huang, Ding, Zhang, Hou, Yao, and Li]{Huang}
Huang, X.H.; Ding, J.F.; Zhang, G.Q.; Hou, Y.; Yao, Y.P.; Li, X.G.
\newblock Size-dependent exchange bias in La$_{0.25}$Ca$_{0.75}$MnO$_3$
  nanoparticles.
\newblock {\em Phys. Rev. B} {\bf 2008}, {\em 78},~224408.
\newblock {\url{https://doi.org/10.1103/PhysRevB.78.224408}}.

\bibitem[Das \em{et~al.}(2011)Das, Majumdar, and Giri]{Sampad1}
Das, S.; Majumdar, S.; Giri, S.
\newblock Multifunctional properties of CoNi alloy embedded in the SiO$_2$
  host: Role of interparticle interaction.
\newblock {\em J. Solid State Chem.} {\bf 2011}, {\em 184},~2215.
\newblock {\url{https://doi.org/10.1016/j.jssc.2011.06.038}}.

\bibitem[Dimitriadis \em{et~al.}(2015)Dimitriadis, Kechrakos,
  Chubykalo-Fesenko, and Tsiantos]{Dimitriadis}
Dimitriadis, V.; Kechrakos, D.; Chubykalo-Fesenko, O.; Tsiantos, V.
\newblock Shape-dependent exchange bias effect in magnetic nanoparticles with
  core-shell morphology.
\newblock {\em Phys. Rev. B} {\bf 2015}, {\em 92},~064420.
\newblock {\url{https://doi.org/10.1103/PhysRevB.92.064420g}}.

\bibitem[Obaidat \em{et~al.}(2017)Obaidat, Nayek, Manna, Bhattacharjee,
  Al-Omari, and Gismelseed]{Obaidat}
Obaidat, I.M.; Nayek, C.; Manna, K.; Bhattacharjee, G.; Al-Omari, I.A.;
  Gismelseed, A.
\newblock Investigating Exchange Bias and Coercivity in
  Fe$_3$O$_4$-$\gamma$-Fe$_2$O$_3$ Core–Shell Nanoparticles of Fixed Core
  Diameter and Variable Shell Thicknesses.
\newblock {\em Nanomaterials} {\bf 2017}, {\em 7},~415.
\newblock {\url{https://doi.org/10.3390/nano7120415}}.

\bibitem[Giri \em{et~al.}(2001)Giri, Ganguli, and Bhattacharya]{GIRI}
Giri, S.; Ganguli, S.; Bhattacharya, M.
\newblock Surface oxidation of iron nanoparticles.
\newblock {\em Appl. Surf. Sci.} {\bf 2001}, {\em 182},~345.
\newblock {\url{https://doi.org/10.1016/S0169-4332(01)00446-9}}.

\bibitem[Zhang \em{et~al.}(2011)Zhang, Hu, Zhao, Tian, Zou, and Xia]{Zhang}
Zhang, L.; Hu, P.; Zhao, X.; Tian, R.; Zou, R.; Xia, D.
\newblock Controllable synthesis of core–shell Co@CoO nanocomposites with a
  superior performance as an anode material for lithium-ion batteries.
\newblock {\em J. Mater. Chem.} {\bf 2011}, {\em 21},~18279.
\newblock {\url{https://doi.org/10.1039/C1JM12990B}}.

\bibitem[Gonz$\acute{a}$lez \em{et~al.}(2017)Gonz$\acute{a}$lez,
  Andr$\acute{e}$s, Ant$\acute{o}$n, De~Toro, Normile, Mu$\tilde{n}$iz,
  Riveiro, and Nogu$\acute{e}$s]{Gonzalez}
Gonz$\acute{a}$lez, J.A.; Andr$\acute{e}$s, J.P.; Ant$\acute{o}$n, R.L.;
  De~Toro, J.A.; Normile, P.S.; Mu$\tilde{n}$iz, P.; Riveiro, J.M.;
  Nogu$\acute{e}$s, J.
\newblock Maximizing exchange-bias in Co/CoO core/shell nanoparticles by
  lattice matching between the shell and the embedding matrix.
\newblock {\em Chem. Mater.} {\bf 2017}, {\em 29},~5200.
\newblock {\url{https://doi.org/10.1021/acs.chemmater.7b00868}}.

\bibitem[Simeonidis \em{et~al.}(2011)Simeonidis, Martinez-Boubeta, Iglesias,
  Cabot, Angelakeris, Mourdikoudis, Tsiaoussis, Delimitis, Dendrinou-Samara,
  and Kalogirou]{Simeonidis}
Simeonidis, K.; Martinez-Boubeta, C.; Iglesias, {\`{O}}.; Cabot, A.;
  Angelakeris, M.; Mourdikoudis, S.; Tsiaoussis, I.; Delimitis, A.;
  Dendrinou-Samara, C.; Kalogirou, O.
\newblock Morphology influence on nanoscale magnetism of Co nanoparticles:
  Experimental and theoretical aspects of exchange bias.
\newblock {\em Phys. Rev. B} {\bf 2011}, {\em 84},~144430.
\newblock {\url{https://doi.org/10.1103/PhysRevB.84.144430}}.

\bibitem[Tracy \em{et~al.}(2005)Tracy, Weiss, Dinega, and Bawendi]{Tracy}
Tracy, J.B.; Weiss, D.N.; Dinega, D.P.; Bawendi, M.G.
\newblock Exchange biasing and magnetic properties of partially and fully
  oxidized colloidal cobalt nanoparticles.
\newblock {\em Phys. Rev. B} {\bf 2005}, {\em 72},~064404.
\newblock {\url{https://doi.org/10.1103/PhysRevB.72.064404}}.

\bibitem[Kovylina \em{et~al.}(2009)Kovylina, Muro, Konstantinovi{\'{c}},
  Varela, Iglesias, Labarta, and Batlle]{Iglesias2}
Kovylina, M.; Muro, M.G.D.; Konstantinovi{\'{c}}, Z.; Varela, V.; Iglesias,
  {\`{O}}.; Labarta, A.; Batlle, X.
\newblock Controlling exchange bias in Co–CoO$_x$ nanoparticles by oxygen
  content.
\newblock {\em Nanotechnology} {\bf 2009}, {\em 20},~175702.
\newblock {\url{https://doi.org/10.1088/0957-4484/20/17/175702}}.

\bibitem[Nogu\'es \em{et~al.}(2006)Nogu\'es, Skumryev, Sort, Stoyanov, and
  Givord]{Nogues2006}
Nogu\'es, J.; Skumryev, V.; Sort, J.; Stoyanov, S.; Givord, D.
\newblock Shell-Driven Magnetic Stability in Core-Shell Nanoparticles.
\newblock {\em Phys. Rev. Lett.} {\bf 2006}, {\em 97},~157203.
\newblock {\url{https://doi.org/10.1103/PhysRevLett.97.157203}}.

\bibitem[Bean and Livingston(1959)]{Bean}
Bean, C.P.; Livingston, J.D.
\newblock Superparamagnetism.
\newblock {\em J. Appl. Phys.} {\bf 1959}, {\em 30},~S120.
\newblock {\url{https://doi.org/10.1063/1.2185850}}.

\bibitem[Bedanta and Kleemann(2008)]{Bedanta}
Bedanta, S.; Kleemann, W.
\newblock Supermagnetism.
\newblock {\em J. Phys. D: Appl. Phys.} {\bf 2008}, {\em 42},~013001.
\newblock {\url{https://doi.org/10.1088/0022-3727/42/1/013001}}.

\bibitem[Chen \em{et~al.}(2005)Chen, Bedanta, Petracic, Kleemann, Sahoo,
  Cardoso, and Freitas]{Bedanta1}
Chen, X.; Bedanta, S.; Petracic, O.; Kleemann, W.; Sahoo, S.; Cardoso, S.;
  Freitas, P.P.
\newblock Superparamagnetism versus superspin glass behavior in dilute magnetic
  nanoparticle systems.
\newblock {\em Phys. Rev. B} {\bf 2005}, {\em 72},~214436.
\newblock {\url{https://doi.org/10.1103/PhysRevB.72.214436}}.

\bibitem[Bedanta \em{et~al.}(2007)Bedanta, Eim\"uller, Kleemann, Rhensius,
  Stromberg, Amaladass, Cardoso, and Freitas]{Bedanta2}
Bedanta, S.; Eim\"uller, T.; Kleemann, W.; Rhensius, J.; Stromberg, F.;
  Amaladass, E.; Cardoso, S.; Freitas, P.P.
\newblock Overcoming the Dipolar Disorder in Dense CoFe Nanoparticle Ensembles:
  Superferromagnetism.
\newblock {\em Phys. Rev. Lett.} {\bf 2007}, {\em 98},~176601.
\newblock {\url{https://doi.org/10.1103/PhysRevLett.98.176601}}.

\bibitem[Luo \em{et~al.}(1991)Luo, Nagel, Rosenbaum, and Rosensweig]{Luo}
Luo, W.; Nagel, S.R.; Rosenbaum, T.F.; Rosensweig, R.E.
\newblock Dipole Interactions with Random Anisotropy in a Frozen Ferrofluid.
\newblock {\em Phys. Rev. Lett.} {\bf 1991}, {\em 67},~2721.
\newblock {\url{https://doi.org/10.1103/PhysRevLett.67.2721}}.

\bibitem[Moscoso-Londo$\tilde{n}$o \em{et~al.}(2017)Moscoso-Londo$\tilde{n}$o,
  Tancredi, Muraca, Z$\acute{e}$lis, Coral, Fern$\acute{a}$ndez~van Raap,
  Wolff, Neu, Damm, de~Oliveira, Pirota, Knobel, and Socolovsky]{Londono}
Moscoso-Londo$\tilde{n}$o, O.; Tancredi, P.; Muraca, D.; Z$\acute{e}$lis, P.M.;
  Coral, D.; Fern$\acute{a}$ndez~van Raap, M.B.; Wolff, U.; Neu, V.; Damm, C.;
  de~Oliveira, C.L.P.;  et~al.
\newblock Different approaches to analyze the dipolar interaction effects on
  diluted and concentrated granular superparamagnetic systems.
\newblock {\em J. Magn. Magn. Mater.} {\bf 2017}, {\em 428},~105.
\newblock {\url{https://doi.org/10.1016/j.jmmm.2016.12.019}}.

\bibitem[Vestal \em{et~al.}(2004)Vestal, Song, and Zhang]{Vestal}
Vestal, C.R.; Song, Q.; Zhang, Z.J.
\newblock Effects of Interparticle Interactions upon the Magnetic Properties of
  CoFe$_2$O$_4$ and MnFe$_2$O$_4$ Nanocrystals.
\newblock {\em J. Phys. Chem. B} {\bf 2004}, {\em 108},~18222.
\newblock {\url{https://doi.org/10.1021/jp0464526}}.

\bibitem[De \em{et~al.}(2005)De, Ray, Panda, Giri, Nakamura, and Kohara]{De}
De, K.; Ray, R.; Panda, R.N.; Giri, S.; Nakamura, H.; Kohara, T.
\newblock The effect of Fe substitution on magnetic and transport properties of
  LaMnO$_3$.
\newblock {\em J. Magn. Magn. Mater.} {\bf 2005}, {\em 288},~339.
\newblock {\url{https://doi.org/10.1016/j.jmmm.2004.09.118}}.

\bibitem[Betancourt-Cantera \em{et~al.}(2019)Betancourt-Cantera,
  S\'anchez-De~Jes\'us, Bolar\'in-Mir\'o, Torres-Villase\~nor, and
  Betancourt-Cantera]{Cantera}
Betancourt-Cantera, J.A.; S\'anchez-De~Jes\'us, F.; Bolar\'in-Mir\'o, A.M.;
  Torres-Villase\~nor, G.; Betancourt-Cantera, L.G.
\newblock Magnetic properties and crystal structure of elemental cobalt powder
  modified by high-energy ball milling.
\newblock {\em J. Mater. Res. Technol.} {\bf 2019}, {\em 8},~4995.
\newblock {\url{https://doi.org/10.1016/j.jmrt.2019.07.048}}.

\bibitem[Rom\'an~de Alba \em{et~al.}(2016)Rom\'an~de Alba, Mart\'inez,
  Guerrero, and Ortega-Zarzosa]{DeAlba}
Rom\'an~de Alba, J.; Mart\'inez, J.R.; Guerrero, A.L.; Ortega-Zarzosa, G.
\newblock Effect of the Silica Cover on the Properties of Co$_3$O$_4$
  Nanoparticles.
\newblock {\em Journal of Superconductivity and Novel Magnetism} {\bf 2016},
  {\em 29},~2651.
\newblock {\url{https://doi.org/10.1007/s10948-016-3595-y}}.

\bibitem[De~Toro \em{et~al.}(2006)De~Toro, Andr\'es, Gonz\'alez, Mu\~niz,
  Mu\~noz, Normile, and Riveiro]{DeToro}
De~Toro, J.A.; Andr\'es, J.P.; Gonz\'alez, J.A.; Mu\~niz, P.; Mu\~noz, T.;
  Normile, P.S.; Riveiro, J.M.
\newblock Exchange bias and nanoparticle magnetic stability in Co-CoO
  composites.
\newblock {\em Phys. Rev. B} {\bf 2006}, {\em 73},~094449.
\newblock {\url{https://doi.org/10.1103/PhysRevB.73.094449c}}.

\bibitem[Williamson and Hall(1953)]{williamson}
Williamson, G.K.; Hall, W.H.
\newblock X-ray line broadening from filed aluminium and wolfram.
\newblock {\em Acta Metall.} {\bf 1953}, {\em 1},~22.
\newblock {\url{https://doi.org/10.1016/0001-6160(53)90006-6}}.

\bibitem[De \em{et~al.}(2015)De, Majumdar, and Giri]{De2015}
De, D.; Majumdar, S.; Giri, S.
\newblock Spin-glass like behaviour in strongly interacting nanocrystalline Ni
  embedded in SiO$_2$.
\newblock {\em J. Magn. Magn. Mater.} {\bf 2015}, {\em 394},~448.
\newblock {\url{https://doi.org/10.1016/j.jmmm.2015.06.071}}.

\bibitem[De \em{et~al.}(2012)De, Karmakar, Bhunia, Bhaumik, Majumdar, and
  Giri]{DEJAP}
De, D.; Karmakar, A.; Bhunia, M.K.; Bhaumik, A.; Majumdar, S.; Giri, S.
\newblock Memory effects in superparamagnetic and nanocrystalline
  Fe$_{50}$Ni$_{50}$ alloy.
\newblock {\em J. Appl. Phys.} {\bf 2012}, {\em 111},~033919.
\newblock {\url{https://doi.org/10.1063/1.3684624}}.

\bibitem[Chattopadhyay \em{et~al.}(2012)Chattopadhyay, Giri, and
  Majumdar]{Chattopadhyay1}
Chattopadhyay, S.; Giri, S.; Majumdar, S.
\newblock Magnetic behaviour of doped dimer compounds
  Sr$_3$Cr$_{2-x}$M$_x$O$_8$ (M = V, Mn).
\newblock {\em Eur. Phys. J. B} {\bf 2012}, {\em 85},~4.
\newblock {\url{https://doi.org/10.1140/epjb/e2011-20660-5}}.

\bibitem[Chattopadhyay \em{et~al.}(2011)Chattopadhyay, Giri, and
  Majumdar]{Chattopadhyay}
Chattopadhyay, S.; Giri, S.; Majumdar, S.
\newblock Broken chain effect in doped SrCuO$_2$.
\newblock {\em J. Phys.: Condens. Matter} {\bf 2011}, {\em 23},~216006.
\newblock {\url{https://doi.org/10.1088/0953-8984/23/21/216006}}.

\bibitem[Sirker \em{et~al.}(2007)Sirker, Laflorencie, Fujimoto, Eggert, and
  Affleck]{Sirker}
Sirker, J.; Laflorencie, N.; Fujimoto, S.; Eggert, S.; Affleck, I.
\newblock Chain Breaks and the Susceptibility of
  Sr2Cu$_{1-x}$Pd$_x$O$_{3+\delta}$ and Other Doped Quasi-One-Dimensional
  Antiferromagnets.
\newblock {\em Phys. Rev. Lett.} {\bf 2007}, {\em 98},~137205.
\newblock {\url{https://doi.org/10.1103/PhysRevLett.98.137205}}.

\bibitem[Iglesias and Kachkachi(2021)]{IglesiasKachkachi}
Iglesias, {\`{O}}.; Kachkachi, H.
\newblock {Single Nanomagnet Behaviour: Surface and Finite-Size Effects}. In
  {\em {New Trends in Nanoparticle Magnetism}}; Springer International
  Publishing: Cham, Switzerland,  2021; pp. 3--38.
\newblock {\url{https://doi.org/10.1007/978-3-030-60473-8_1}}.

\bibitem[Sánchez \em{et~al.}(2020)Sánchez, Vasilakaki, Lee, Normile, Muscas,
  Murgia, Andersson, Singh, Mathieu, Nordblad, Ricci, Peddis, Trohidou,
  Nogués, and De~Toro]{Sanchez}
Sánchez, E.H.; Vasilakaki, M.; Lee, S.S.; Normile, P.S.; Muscas, G.; Murgia,
  M.; Andersson, M.S.; Singh, G.; Mathieu, R.; Nordblad, P.;  et~al.
\newblock Simultaneous Individual and Dipolar Collective Properties in Binary
  Assemblies of Magnetic Nanoparticles.
\newblock {\em Chemistry of Materials} {\bf 2020}, {\em 32},~969.
\newblock {\url{https://doi.org/10.1021/acs.chemmater.9b03268}}.

\bibitem[Sánchez \em{et~al.}(2022)Sánchez, Vasilakaki, Lee, Normile,
  Andersson, Mathieu, López-Ortega, Pichon, Peddis, Binns, Nordblad, Trohidou,
  Nogués, and De~Toro]{Sanchez2}
Sánchez, E.H.; Vasilakaki, M.; Lee, S.S.; Normile, P.S.; Andersson, M.S.;
  Mathieu, R.; López-Ortega, A.; Pichon, B.P.; Peddis, D.; Binns, C.;  et~al.
\newblock Crossover From Individual to Collective Magnetism in Dense
  Nanoparticle Systems: Local Anisotropy Versus Dipolar Interactions.
\newblock {\em Small} {\bf 2022}, {\em 18},~2106762.
\newblock {\url{https://doi.org/10.1002/smll.202106762}}.

\bibitem[Iglesias \em{et~al.}(2008)Iglesias, Labarta, and Batlle]{Iglesias}
Iglesias, {\`{O}}.; Labarta, A.; Batlle, X.
\newblock Exchange Bias Phenomenology and Models of Core/Shell Nanoparticles.
\newblock {\em J. Nanosci. Nanotechnol.} {\bf 2008}, {\em 8},~2761.
\newblock {\url{https://doi.org/10.1166/jnn.2008.18306}}.

\bibitem[Kostopoulou \em{et~al.}(2014)Kostopoulou, Brintakis, Vasilakaki,
  Trohidou, Douvalis, Lascialfari, Manna, and Lappas]{Kostopoulou2014}
Kostopoulou, A.; Brintakis, K.; Vasilakaki, M.; Trohidou, K.N.; Douvalis, A.P.;
  Lascialfari, A.; Manna, L.; Lappas, A.
\newblock Assembly-mediated interplay of dipolar interactions and surface spin
  disorder in colloidal maghemite nanoclusters.
\newblock {\em Nanoscale} {\bf 2014}, {\em 6},~3764--3776.
\newblock {\url{https://doi.org/10.1039/C3NR06103E}}.

\bibitem[Silva \em{et~al.}(2022)Silva, Vasilakaki, Cabreira~Gomes, Aquino,
  Campos, Dubois, Perzynski, Depeyrot, and Trohidou]{Silva2022}
Silva, F.G.d.; Vasilakaki, M.; Cabreira~Gomes, R.; Aquino, R.; Campos, A.F.C.;
  Dubois, E.; Perzynski, R.; Depeyrot, J.; Trohidou, K.
\newblock {A numerical study on the interplay between the intra-particle and
  interparticle characteristics in bimagnetic soft/soft and hard/soft
  ultrasmall nanoparticle assemblies}.
\newblock {\em Nanoscale Adv.} {\bf 2022}.
\newblock {\url{https://doi.org/10.1039/D1NA00894C}}.

\bibitem[Omelyanchik \em{et~al.}(2021)Omelyanchik, Villa, Vasilakaki, Singh,
  Ferretti, Ponti, Canepa, Margaris, Trohidou, and Peddis]{Omelyanchik2021}
Omelyanchik, A.; Villa, S.; Vasilakaki, M.; Singh, G.; Ferretti, A.M.; Ponti,
  A.; Canepa, F.; Margaris, G.; Trohidou, K.N.; Peddis, D.
\newblock Interplay between inter- and intraparticle interactions in
  bi-magnetic core/shell nanoparticles.
\newblock {\em Nanoscale Adv.} {\bf 2021}, {\em 3},~6912--6924.
\newblock {\url{https://doi.org/10.1039/D1NA00312G}}.

\bibitem[Iglesias \em{et~al.}(2005)Iglesias, Batlle, and Labarta]{Iglesias1}
Iglesias, {\`{O}}.; Batlle, X.; Labarta, A.
\newblock Microscopic origin of exchange bias in core/shell nanoparticles.
\newblock {\em Phys. Rev. B} {\bf 2005}, {\em 72},~212401.
\newblock {\url{https://doi.org/10.1103/PhysRevB.72.212401}}.

\bibitem[Torres \em{et~al.}(2017)Torres, Morales, Schuller, and
  Kiwi]{Torres2017}
Torres, F.; Morales, R.; Schuller, I.K.; Kiwi, M.
\newblock Dipole-induced exchange bias.
\newblock {\em Nanoscale} {\bf 2017}, pp. {17074--17079}.
\newblock {\url{https://doi.org/10.1039/C7NR05491B}}.

\bibitem[Moya \em{et~al.}(2015)Moya, Iglesias, Batlle, and Labarta]{Moya}
Moya, C.; Iglesias, {\`{O}}.; Batlle, X.; Labarta, A.
\newblock Quantification of Dipolar Interactions in Fe3–xO4 Nanoparticles.
\newblock {\em The Journal of Physical Chemistry C} {\bf 2015}, {\em
  119},~24142.
\newblock {\url{https://doi.org/10.1021/acs.jpcc.5b07516}}.

\bibitem[Eftaxias and Trohidou(2005)]{Eftaxias}
Eftaxias, E.; Trohidou, K.N.
\newblock Numerical study of the exchange bias effects in magnetic
  nanoparticles with core/shell morphology.
\newblock {\em Phys. Rev. B} {\bf 2005}, {\em 71},~134406.
\newblock {\url{https://doi.org/10.1103/PhysRevB.71.134406}}.

\bibitem[Sort \em{et~al.}(2004)Sort, Langlais, Doppiu, Dieny, Suri{\~{n}}ach,
  Mu{\~{n}}oz, Bar{\'{o}}, Laurent, and Nogu{\'{e}}s]{Sort}
Sort, J.; Langlais, V.; Doppiu, S.; Dieny, B.; Suri{\~{n}}ach, S.; Mu{\~{n}}oz,
  J.S.; Bar{\'{o}}, M.D.; Laurent, C.; Nogu{\'{e}}s, J.
\newblock Exchange bias effects in Fe nanoparticles embedded in an
  antiferromagnetic Cr$_2$O$_3$ matrix.
\newblock {\em Nanotechnology} {\bf 2004}, {\em 15},~S211.
\newblock {\url{https://doi.org/10.1088/0957-4484/15/4/017}}.

\end{thebibliography}
\end{adjustwidth}
\end{document}